\documentclass[showpacs,preprintnumbers,amsmath,amssymb]{revtex4}

\usepackage{graphicx}

\begin{document}


\title{Natural Metric for Quantum Information Theory}

\author{P.W. Lamberti$^{1}$, M. Portesi$^{2}$ and J. Sparacino$^{1}$} \affiliation{$^1$
Facultad de Matem\'atica, Astronom\'{\i}a y F\'{\i}sica,
Universidad Nacional de C\'ordoba  and CONICET, Ciudad
Universitaria, 5000 C\'ordoba, Argentina
\\\\ $^2$ Instituto de F\'{i}sica La Plata (CONICET) \ and \ Departamento de F\'{i}sica, Facultad de Ciencias Exactas,
Universidad Nacional de La Plata, CC 67, 1900 La Plata, Argentina} %

\date{\today}
\begin{abstract} %
We study in detail a very natural metric for quantum states. This new
proposal has two basic ingredients: entropy and purification. The
metric for two mixed states is defined as the square root of the entropy of the average
of representative purifications of those states.
Some basic properties are analyzed and its relation with other
distances is investigated. As an illustrative application, the proposed metric is evaluated for
1-qubit mixed states.
\end{abstract}

\pacs{03.67.$-$a; 03.67.Mn; 03.65.$-$w}

\maketitle

\section{\label{sec:intro}Introduction}
The study of distances and metrics between quantum states is a
topic of permanent interest, which has been lately rekindled on
account of problems emerging in quantum information theory (QIT)
\cite{Nielsen}. Distances are used as measure of
distinguishability between quantum states \cite{Vedral} and in the
definition of the degree of entanglement \cite{Vedral97}, just to
mention two very relevant examples. They also  characterize the
geometrical structure of the space of quantum states \cite{BZ06}.

In the mathematical formalism of quantum mechanics the states of a
physical system $\cal{S}$ are represented by operators (density
operators) acting on a Hilbert space $\cal{H}$. More precisely the
states of the system $\cal{S}$ are represented by the elements of
${\cal{B}}({\cal{H}})_1^+$, that is, the set of positive, trace-one
operators on $\cal{H}$. The notion of a state as a unit vector of
$\cal{H}$ refers to the extremal elements of
${\cal{B}}({\cal{H}})_1^+$ ($\rho \in
{\cal{B}}({\cal{H}})_1^+$ is extremal if and only if it is
idempotent, $\rho^2 = \rho$). In this case $\rho$ is of the form
$|\psi\rangle\langle\psi|$ for some unit vector
$|\psi\rangle \in \cal{H}$, and it is called a
pure state. If $\rho$ is not idempotent, the corresponding state
is called mixed.

In introducing distances between quantum states different roads
have been traversed \cite{Wootters, Braunstein, Uhlmann, Jozsa,
Lee, Zhi, Mendo}. For example Wootters arrived at the distance
\begin{equation}
d_W(|\psi\rangle,|\varphi\rangle)  =
\arccos(|\langle\psi|\varphi\rangle|) \label{Wootters}
\end{equation}
by analyzing the statistical fluctuations in the outcomes of
measurements into the quantum mechanics formalism \cite{Wootters}.
In Eq.~(\ref{Wootters}), $\langle\psi|\varphi\rangle$ represents the
inner product between the pure states $|\psi\rangle$ and
$|\varphi\rangle$, and therefore $d_W$ gives the ``angle" between these
two states.

Another way of dealing with the problem of introducing distances
between quantum states is to generalize the notions of  distance
defined in the space of classical probability distributions. This
is the case of the relative entropy, which is a generalization of
information theoretic Kullback--Leibler divergence. The relative
entropy of an operator $\rho$ with respect to an operator
$\sigma$, both belonging to ${\cal{B}}({\cal{H}})_1^+$, is
\begin{equation}
S(\rho,\sigma)= \mathrm{Tr}[\rho(\log_2\rho-\log_2\sigma)]
\label{rel-ent}
\end{equation}
where $\log_2$ stands for logarithm in base two. The relative
entropy is not a metric (because it is not symmetric and does not
verify the triangle inequality). Even worst, it may even be
unbounded. In particular, the relative entropy is well defined
only when the support of $\sigma$ is equal to or larger than that
of $\rho$ \cite{Lindblad} (the support of an operator is the
subspace spanned by the eigenvectors of the operator with nonzero
eigenvalues). This is a strong restriction which is violated in
some physically relevant situations, as for example when $\sigma$
is a pure reference state.

Recently we have investigated a distance between mixed quantum
states that was named the quantum Jensen--Shannon divergence (QJSD)
and that is a symmetrized version of relative entropy
(\ref{rel-ent}):
\begin{equation}
D_{JS}(\rho,\sigma) =
\frac{1}{2}\left[S\left(\rho,\frac{\rho+\sigma}{2}\right)+
S\left(\sigma,\frac{\rho+\sigma}{2}\right)\right] .
\label{JS-relative-entropy}
\end{equation}
The main properties of the QJSD as a distinguishability measure
have been presented in Ref.~\cite{Majtey}; the metric character of the
square root of $D_{JS}$ has also been reported recently
\cite{last, Topsoe}. It has several interesting interpretations
into the realm of QIT and it has been applied as a measure of the
degree of entanglement \cite{Erice}.

In this work we  propose an alternative metric between quantum
states, that we think, from a conceptual point of view, is a very
natural one and which is, in some sense, a derivative of the QJSD.
This proposal hinges on two central concepts of QIT: entropy and
purification.
Relevant properties of these notions are quoted in the following
section. The rest of the paper is organized as follows: in
Section~\ref{sec:pure} we consider a measure of distance for pure
states defined using von Neumann entropy and we study its
properties. Section~\ref{sec:mixed} is devoted to the new proposal
of a distance between mixed states by recourse to the concept of
purification. We then present in Section~\ref{sec:applic} an
application for the case of 1-qubit mixed states with a numerical
example. Finally, concluding remarks are drawn in
Section~\ref{sec:conclu}.

\section{Some comments on Entropy and Purification} %
\label{sec:keyconcepts} %

\textit{Entropy} is a fundamental notion in classical and in
quantum information theory. In the classical case most of the
results concerning the coding problem can be expressed in terms
of the Shannon entropy
\begin{equation}
H_S(P) = - \sum_i p_i \log_2 p_i \label{Shannon}
\end{equation}
where $P=\{p_i \geq 0, \sum_i p_i=1 \}$ is a (discrete)
probability distribution \cite{Thomas}. In the quantum context the
expression for entropy differs from the Shannon entropy. For a
mixed state described  by the density operator $\rho$, von Neumann
defined the entropy of $\rho$ as
\begin{equation}
H_N(\rho) = - \mathrm{Tr}(\rho \log_2 \rho)= -\sum_i \lambda_i
\log_2 \lambda_i \label{vneu}
\end{equation}
with $\{\lambda_i\}$ being the set of eigenvalues of the operator $\rho$.

The von Neumann entropy has several interesting properties
\cite{Wehrl}. Among them we remark the one that gives an upper
bound for a convex combination $\rho=\sum_i p_i \rho_i$ \
(with $\sum_i p_i =1$):
\begin{equation}
H_N(\rho) \leq \sum_i p_i H_N(\rho_i) + H_S(\{p_i\}) . \label{ineqvn}
\end{equation}
Equality is verified if and only if the states $\rho_i$ have
supports in orthogonal subspaces.

Although expressions (\ref{Shannon}) and (\ref{vneu}) look
similar, they are quite different. These differences are
particularly relevant for quantum information theory. For example,
if $a$ is a message taken from the source $\cal{A}$ and $p(a)$ is
the probability of the message $a$, the Shannon entropy of the
source is
\[
H_S[{\cal{A}}]=-\sum_a p(a) \log_2 p(a) .
\]
Let us now suppose that we have a quantum signal source, that is a
device that codes a message $a$ taken from the source $\cal{A}$
into a signal state $|a_S\rangle$ of a quantum system $\cal{S}$.
The ensemble of signals from the signal source will be represented
by the density operator
\begin{equation}
\mu = \sum_a p(a) |a_S \rangle \langle a_S| . \label{pii}
\end{equation}
Then, if the signal states $|a_S\rangle$ are not orthogonal and
from (\ref{ineqvn}), the inequality
\begin{equation}
H_N(\mu) < H_S(\{p(a)\}) \label{ineq}
\end{equation}
is satisfied. The  physical consequences of this inequality have
been analyzed in detail by Jozsa and Schlienz \cite{Jozsa2}.

Another important property of von Neumann entropy is that it gives
the number of qubits necessary to represent a quantum signal
faithfully. Indeed, let us suppose that Alice has a source of pure
qubit signal states $|\psi\rangle$ and $|\varphi\rangle$. Each
emission is chosen to be $|\psi\rangle$ or $|\varphi\rangle$ with
an equal prior probability one half. Then in this case the density
matrix of the source  is $\pi = \frac{1}{2} \left(|\psi\rangle
\langle \psi| + |\varphi\rangle \langle\varphi| \right)$. Alice
may  communicate the sequence of states to Bob by transmitting one
qubit per emitted state. But, according to the quantum source
coding theorem, the quantity
\[
H_N\left(\frac{|\psi\rangle\langle\psi|+|\varphi\rangle\langle\varphi|}{2}\right)
\]
gives the lowest number of qubits per state that Alice needs to
communicate the quantum information (with arbitrarily high
fidelity) \cite{Schumacher}.

\hfill %

\textit{Purification} is the second key concept we use in proposing a metric for %
quantum states. It has to do
with the fact that every mixed quantum state can be interpreted as
being part of a higher-dimensional pure state. From a physical
point of view the notion of purification provides, for example,
insight into the mechanism of quantum decoherence \cite{Zurek}.
Formally it can be described in the following way: let $\rho$ be
any mixed state on the Hilbert space $\cal{H}$. A purification of
$\rho$ is any pure state $|\psi\rangle$ in any extended Hilbert
space ${\cal{H}}\otimes{\cal{H}}_{aux}$ with the property that
\[
\rho=\mathrm{Tr}_{aux} |\psi\rangle\langle\psi|
\]
where $\mathrm{Tr}_{aux}$ stands for the partial trace on the
Hilbert space ${\cal{H}}_{aux}$. In other words a purification is
any pure state having $\rho$ as the reduced state for subsystem
\cite{Jozsa}.

If $\rho$ admits the decomposition $\rho= \sum_i p_i \,
|e_i\rangle\langle e_i|$ where $p_i \geq 0$ and $\sum_i p_i=1$, an
example for a purification of $\rho$ is given by
\begin{equation}
|\psi\rangle = \sum_i \sqrt{p_i} \, |e_i\rangle \otimes |a_i\rangle
\end{equation}
with the auxiliary states $|a_i\rangle$ being mutually orthogonal.

We complete this section by remarking two facts on purifications that %
will be used later on:
\begin{enumerate}
\item It can be shown that for two purifications $|\psi_1\rangle$ and $|\psi_2\rangle$ of the state
$\rho$, there exists a unitary transformation $U$ acting on
${\cal{H}}_{aux}$ such that
\begin{equation}
|\psi_1\rangle = (I \otimes U) |\psi_2\rangle \label{unitary}
\end{equation}
where $I$ is the identity operator on the space $\cal{H}$
\cite{Nielsen}.
\item If one is interested in purifications of
two states it can be assumed, without loss of generality, that the
purifications lie in the same extended Hilbert space \cite{Jozsa}.
\end{enumerate}

\section{von Neumann entropy as a metric for pure states} %
\label{sec:pure}

Let $|\psi\rangle$ and $|\varphi\rangle$ be two pure states in a
given Hilbert space $\cal{H}$. We define the distance between
these two states in the form:
\begin{equation}
D_N(|\psi\rangle,|\varphi\rangle) \equiv
\sqrt{H_N\left(\frac{|\psi\rangle\langle\psi|+|\varphi\rangle\langle\varphi|}{2}\right)} .
\label{pure}
\end{equation}
It should be noted that this expression corresponds to the square
root of the QJSD when evaluated between pure states. Indeed, in
terms of the von Neumann entropy (\ref{vneu}), the QJSD
(\ref{JS-relative-entropy}) can be rewritten in the form:
\[
D_{JS}(\rho,\sigma)= H_N\left(\frac{\rho + \sigma}{2}\right) - \frac{1}{2}
H_N(\rho) - \frac{1}{2} H_N(\sigma) .
\]
But for a pure state the von Neumann entropy vanishes; then for
the corresponding operators $\rho=|\psi\rangle\langle\psi|$ and
$\sigma=|\varphi\rangle\langle \varphi|$ it results:
\[
D_{JS}(|\psi\rangle\langle\psi|, |\varphi\rangle\langle \varphi|)=
D_N^2(|\psi\rangle,|\varphi\rangle) .
\]

After some algebra, we can rewrite the square of Eq.~(\ref{pure}),
in
terms of the inner product between $|\psi\rangle$ and $|\varphi\rangle$:
\begin{equation}
D_N^2(|\psi\rangle,|\varphi\rangle) = \Phi(|\langle \psi |
\varphi \rangle|)
\label{entro2}
\end{equation}
where
\begin{equation}
\Phi(x) \equiv -\left(
\frac{1-x}{2}\right)
\log_2\left(\frac{1-x}{2}\right)
-\left( \frac{1+x}{2}\right)
\log_2\left(\frac{1+x}{2}\right) \ , \qquad 0\leq x \leq 1 .
\label{Phi}
\end{equation}

The main properties of $D_N(|\psi\rangle,|\varphi\rangle)$
given by Eq.~(\ref{pure}) are:
\begin{itemize}
\item It is symmetric.
\item It vanishes if and only if $|\psi\rangle = e^{i \alpha}
|\varphi\rangle$ \ (i.e., both states belong to the same ray).
\item It is bounded: $0 \leq D_N(|\psi\rangle,|\varphi\rangle) \leq 1$ \ (this can be easily verified from inequality~(\ref{ineqvn}) and
from the fact that the von Neumann entropy vanishes for a pure
state).
\item It verifies the triangle inequality \ (but $H_N\left(\frac{|\psi\rangle\langle\psi|+
|\varphi\rangle\langle\varphi|}{2}\right)$ does not) \cite{last,
Topsoe}.
\end{itemize}
All these properties give $D_N$ the metric character.

The distance $D_N$ between two neighboring pure states
$|\psi\rangle=\sum_j \sqrt{p_j}\, e^{i \phi_j} |j\rangle$ and
$|\tilde{\psi}\rangle = |\psi\rangle + |d \psi\rangle= \sum_j
\sqrt{p_j + dp_j} \, e^{i(\phi_j + d\phi_j)} |j\rangle$, where
$\{|j\rangle\}$ in an orthonormal basis, is given by:
\begin{equation}
D_N^2(|\psi\rangle,|\tilde{\psi}\rangle) \simeq \frac{1}{8} \sum_j
\frac{dp_j^2}{p_j} .
\end{equation}

From expression~(\ref{Wootters}) it is easily checked that, up to
second order in $dp_j$, the following relation exists between
Wootters distance and the distance $D_N$ given in (\ref{pure}):
\[
d_W^2(|\psi\rangle, |\tilde{\psi}\rangle)= 2 \,
D_N(|\psi\rangle,|\tilde{\psi}\rangle) .
\]

A similar relation is also true for the Fubini--Study metric
\cite{nos1}.

\section{Extension of the definition of the metric to mixed states} %
\label{sec:mixed}

In the previous section we have thought the square root of the von %
Neumann entropy of the average $\frac 12 \left( |\psi\rangle \langle %
\psi| + |\varphi\rangle \langle\varphi| \right)$ as a true
metric between two pure states $|\psi\rangle$ and
$|\varphi\rangle$. Here we investigate its extension to mixed
states.

We start by recalling that the Bures distance between two mixed
states, that is for two operators $\rho$ and $\sigma$
belonging to ${\cal{B}}({\cal{H}})_1^+$, is given by
\[
D_B(\rho,\sigma) = \sqrt{2-2\,F(\rho,\sigma)}
\]
where $F(\rho,\sigma)= \mathrm{Tr} \sqrt{ \rho^{1/2}\,\sigma
\rho^{1/2} }$ \, is the fidelity function (see, for instance, Ref.~\cite{Nielsen}). %
A well-known result (Uhlmann's theorem %
\cite{Uhlmann, Jozsa}) asserts that the fidelity $F(\rho,\sigma)$ %
can be expressed in the form
\begin{equation}
F(\rho,\sigma)=
\max_{|\psi\rangle,|\varphi\rangle}|\langle\psi|\varphi\rangle|
\label{jozsa}
\end{equation}
where the maximization is performed over all purifications $|\psi\rangle$ of
$\rho$ and all purifications $|\varphi\rangle$ of $\sigma$. Then,
the Bures metric can be rewritten in the form:
\begin{equation}
D_B(\rho,\sigma) = \min_{|\psi\rangle,|\varphi\rangle}
\sqrt{2-2|\langle\psi|\varphi\rangle|} \label{bures}
\end{equation}
where the minimum is taken over all purifications of $\rho$ and
$\sigma$.

By mimicking this last expression, we can define from
Eq.~(\ref{pure}), a metric for arbitrary mixed states $\rho$
and $\sigma$ belonging to ${\cal{B}}({\cal{H}})_1^+$:
\begin{equation}
D_N(\rho,\sigma) \equiv \min_{|\psi\rangle,|\varphi\rangle}
\sqrt{H_N \left( \frac{|\psi\rangle \langle \psi| +
|\varphi\rangle \langle\varphi|}{2} \right)} \label{new}
\end{equation}
where, once again, the minimum is taken over all purifications
$|\psi\rangle$ of $\rho$ and all purifications $|\varphi\rangle$ of
$\sigma$. Recalling Eq.~(\ref{entro2}) and due to the decreasing nature of $\Phi(x)$ as a function
of $x=|\langle\psi|\varphi\rangle|\,\in[0,1]$, Eq.~(\ref{Phi}),
to seek the minimum out in our proposal~(\ref{new}) is equivalent to look for
the purifications that maximize the overlap
$|\langle\psi|\varphi\rangle|$. Clearly, by construction, the metric character of
$D_N(\rho, \sigma)$ is kept for mixed states. It also is always
well defined.

From a conceptual point of view, our proposal is equivalent to
replace the problem of measuring the distance between two
arbitrary mixed states, $\rho$ and $\sigma$, by the problem of
distinguishability of all the ensembles ${\cal{E}}=
\{|\psi\rangle, |\varphi\rangle; p_1=\frac{1}{2},
p_2=\frac{1}{2}\}$ built from purifications of $\rho$ and $\sigma$
respectively \cite{Jozsa2}. This is the central point in our
proposal.

According to identity~(\ref{jozsa}), the metric $D_N(\rho,\sigma)$
can be expressed in terms of the fidelity in the form:
\begin{equation}
D_N(\rho,\sigma)=\sqrt{\Phi(F(\rho,\sigma))} \label{alter2}
\end{equation}
By using this representation, we can derive the main properties of
the metric $D_N(\rho,\sigma)$. For example
\begin{itemize}
\item $D_N$ is invariant under unitary transformations: \
$D_N(U \rho U^\dag, U \sigma U^\dag) = D_N(\rho, \sigma) $.
\item If ${\cal{Q}}$ is a trace-preserving quantum operation, then: \
$D_N({\cal{Q}}(\rho), {\cal{Q}}(\sigma)) \leq D_N(\rho, \sigma)$.
\end{itemize}
As a textbook exercise we can derive an alternative expression for
fidelity \cite{Nielsen}:
\begin{equation}
F(\rho, \sigma) = \max_{|\varphi\rangle}|\langle\psi|\varphi\rangle|
\label{alternative}
\end{equation}
where $|\psi\rangle$ is any \textit{fixed} purification of $\rho$, and the
maximization is performed over all purifications $|\varphi\rangle$ of $\sigma$. Therefore, by
using this expression and Eq.~(\ref{alter2}), the new metric~(\ref{new})
can be reexpressed in a more convenient way:
\begin{equation}
D_N(\rho,\sigma) = \min_{|\varphi\rangle} \sqrt{H_N \left(
\frac{|\psi\rangle \langle \psi| + |\varphi\rangle
\langle\varphi|}{2} \right)} \label{new1}
\end{equation}
where $|\psi\rangle$ is any fixed purification of $\rho$, and the
minimization is taken over all purifications $|\varphi\rangle$ of $\sigma$.

Several interesting questions arise from the definition of distance between quantum states given by Eq. %
(\ref{new}). Among them we remark the following two:
\begin{itemize}
\item The first one has to do with the computability of
$D_N(\rho,\sigma)$. As it was said, the purifications of $\rho$
and $\sigma$ can be thought as belonging to the same extended
Hilbert space. Furthermore, according to Eq.~(\ref{unitary}),
every purification is related to another one by a unitary
transformation. Therefore the technical task of finding the minimum in Eq.~(\ref{new1}) is %
equivalent to finding unitary operations acting on the auxiliary
Hilbert space that maximize the overlap
$|\langle\psi|\varphi\rangle|$. This point will be analyzed, for a
particular example, in the next section.
\item The second point to be remarked is the relevance of the
metric $D_N(\rho,\sigma)$ to the light of decoherence. Decoherence can be
viewed as the unitary evolution of a compound system consisting of
the system itself and its environment, under which both components
become entangled \cite{Zurek}. In this context, the
auxiliary space ${\cal{H}}_{aux}$ corresponds to the environment.
\end{itemize}

\section{Evaluation of the metric $D_N$ for 1-qubit mixed states}
\label{sec:applic}

As an illustrative application, we now evaluate the metric $D_N$ given in Eq.~(\ref{new}), for two 1-qubit
states. In this case it is convenient to use the Bloch representation for
a mixed 1-qubit state:
\begin{equation}
\rho= \frac{I_2+\vec{r}\cdot\vec{\sigma}}{2} \label{qubit}
\end{equation}
where $I_n$ denotes the $n\times n$ identity matrix, the Bloch
vector $\vec{r}$ is a 3-dimensional vector such that
$\|\vec{r}\|\leq 1$, and the vector $\vec{\sigma}$ has as
components the Pauli matrices. The matrix $\rho$ belongs to the
complex plane $C^2$. In Ref.~\cite{constantinescu} it is shown
that a purification of an arbitrary density $\rho$ acting on $C^2
\otimes C^2$ (that is, a $4 \times 4$ matrix) is given by
\begin{equation}
P_{\rho} = \frac{1}{2} \left(I_4 + \sum_i r_i \,\sigma_i \otimes I_2
+ \sum_i \gamma_i \,I_2 \otimes \sigma_i + \sum_{i,j} A_{ij} \,
\sigma_i \otimes \sigma_j \right) \label{purif}
\end{equation}
where $i,j=1,2,3$, \ $r_i$ are the components of the vector $\vec{r}$,
$\gamma_i$
are the components of the vector $\vec{\gamma}= A^T \vec{r}$, and
the $3 \times 3$ real matrix $A$ is a solution of the system of
equations:
\begin{eqnarray}
A A^T &=& (1 - \|\vec{r}\|) \, I_3 + \vec{r} \, \vec{r}^T \nonumber \\
\det(A) &=& \|\vec{r}\|^2-1
\label{system}
\end{eqnarray}
The result by Constantinescu {\it et al.} \cite{constantinescu} establishes further that all
purification of $\rho$ is of the form (\ref{purif}) with
matrix $A=\tilde{A} S$, with $S$ being an element of the Lie group
$SO(3,\mathbb{R}^3)$ and $\tilde{A}$ a particular solution of the
system (\ref{system}). Therefore, in this case the evaluation of the
minimum in Eq.~(\ref{new1}) to obtain $D_N$ leads to an extremization problem over the
parameter space of the Lie group $SO(3,\mathbb{R}^3)$.

As a particular example, let us show how to find this minimum by evaluating the distance
$D_N(\rho, {\cal{E}}_p(\rho))$ between an input state $\rho$ and
the state emerging from a depolarizing channel
\begin{equation}
{\cal{E}}_p(\rho)=\frac{p\,I_2}{2} + (1-p)\, \rho ,  \qquad 0 \leq p \leq 1 .
\label{depolarizing} \end{equation}
The depolarizing channel can be viewed as a process that takes a system to a fully mixed state with
probability $p$, or leaves it unchanged with probability $1-p$.
The effect of the depolarizing channel (\ref{depolarizing}) over the state $\rho$ is to
``contract'' the vector $\vec{r}$ in the factor $(1-p)$:
$$\rho(\vec r) \rightarrow {\cal E}_p(\rho) = \rho(\vec{r'}) \ ,  \quad \vec{r'}=(1-p)\,\vec r$$.

A strategy for evaluating the metric $D_N(\rho,
{\cal{E}}_p(\rho))$ could be established in the following way: let
$A$ be a solution
of the system (\ref{system}) corresponding to a Bloch vector $\vec{r}$, and let
$A_p$ be the solution of (\ref{system}) corresponding to
$(1-p)\,\vec{r}$. A simple calculation shows that
\begin{eqnarray}
A_p A_p^T & = & A A^T + f(p) \, \Omega  \nonumber \\
\det A_p & = & \det A - f(p) \|\vec{r}\|^2 \label{op2}
\end{eqnarray}
where $f(p) = 1-(1-p)^2$ \ and \ $\Omega=\|\vec{r}\|^2 I_3 -
\vec{r}\, \vec{r}\;^T$. Now we take a fixed purification of $\rho$
characterized by the matrix $A$ (solution of (\ref{system})) and
evaluate a matrix $\tilde{A}_p$ that satisfies (\ref{op2})
(associated with purification $P_{{\cal{E}}_p(\rho)}$). Then every
purification of ${\cal{E}}_p(\rho)$ can be expressed in the form
$A_p = \tilde{A}_p S$ with $S$ being an element of
$SO(3,\mathbb{R}^3)$. Thus to find $S \in SO(3,\mathbb{R}^3)$ that
minimizes our proposal~(\ref{new1}), is equivalent to find $S \in
SO(3,\mathbb{R}^3)$ such that minimizes the norm $\|\tilde{A}_p S
- A\|_{HS}$  \cite{constantinescu}. Here $\|.\|_{HS}$ stands for
the Hilbert--Schmidt norm. This is the classical Procrustes
problem. An algorithmic solution of the Procrustes problem can be
found in Ref.~\cite{Schonemann}. By solving the systems of
equations (\ref{system}) and (\ref{op2}) and applying the
algorithm for solving the Procrustes problem, we can find the
purification $P_{{\cal{E}}_p(\rho)}$ that extremizes (\ref{new1}).
Then by calculating the eigenvalues of the ``average''
$\frac{P_{\rho} + P_{{\cal{E}}_p(\rho)}}{2}$ we can evaluate the
distance $D_N(\rho, {\cal{E}}_p(\rho))$. Figure~1 shows the values
of the new distance $D_N(\rho, {\cal{E}}_p(\rho))$ as a function
of the parameter $p$, corresponding to different 1-qubit states
$\rho$ characterized by different values of the norm of the Bloch
vector $\vec{r}$.

\begin{figure}
\begin{center}
\includegraphics[scale=0.6]{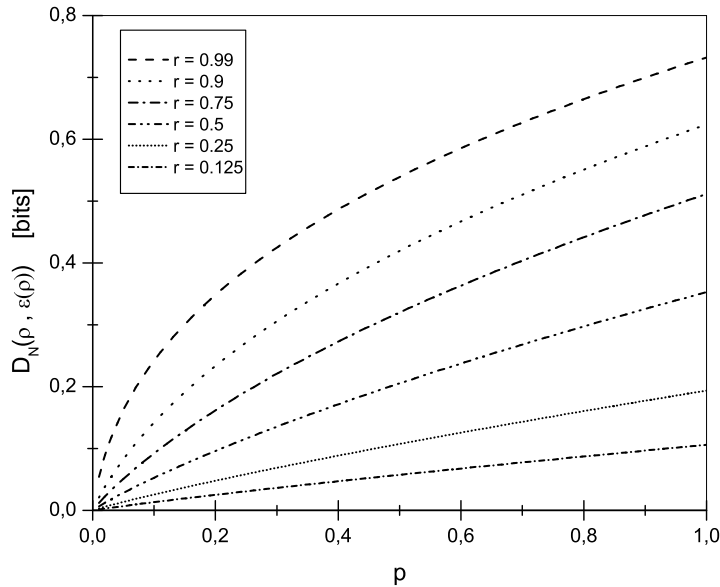}
\vskip -8mm \caption{Values of the distance $D_N(\rho, {\cal{E}}_p(\rho))$ between the state $\rho$ and the emerging
ones from the depolarizing channel as a
function of $p$, for different values of the norm $r=\|\vec{r}\|$ of the Bloch vector
 (see Eqs.~(\ref{qubit}) and (\ref{depolarizing})).}
\end{center}
\label{Fig1}
\end{figure}

\section{Summary and conclusions} %
\label{sec:conclu}

We have studied in detail a new proposal in order to define a metric %
for quantum states. In the case of two pure states
$|\psi\rangle$ and $|\varphi\rangle$, the metric
$D_N(|\psi\rangle,|\varphi\rangle)$ given in Eq.~(\ref{pure}) corresponds to the square root %
of a von Neumann entropy, more precisely the square root of the
quantum Jensen--Shannon divergence. In the case of two mixed
states $\rho$ and $\sigma$, the metric $D_N(\rho,\sigma)$ given in Eq.~(\ref{new}) is %
defined as a natural extension of the latter and results to
be given as the minimum of the distance between pure states
corresponding to purifications of both mixed densities. From a
physical point of view we have replaced the problem of evaluating
the distance between two mixed states by the problem of
distinguishing ensembles built from purifications of that states.

We have analyzed some of the properties of the new proposal,
justifying its metric character, and illustrated how to implement
the minimization procedure with an application for two 1-qubit
mixed states. As a particular example, we have considered a %
depolarizing channel. By using the representation for the
purifications of multiple qubits states \cite{constantinescu}, an
evaluation of the distance $D_N$ in this case can be also
performed.

It should be emphasized that our proposal, although close, differs %
from another ones. For example the definition of the Shannon
distinguishability \cite{Fuchs} involves the Shannon entropy and
quantum \textit{measurement processes}. It should be stressed also
that, although the computational effort of evaluating the minimum
in (\ref{new}) and (\ref{bures}) could be similar, the metric
$D_N$ is grounded in two very relevant physical concepts as are
entropy and purification. The study of some applications of this
metric in the context of decoherence and entanglement is in
progress.


\begin{acknowledgements}

This work was  partially supported by CONICET (National Research Council), Argentina. %
PWL wants to thank SECyT--UNC (Argentina) for financial
assistance. The authors want to thank Prof.\ Domingo Prato for
useful discussion.
\end{acknowledgements}

\end{document}